\documentclass[twocolumn,showpacs,preprintnumbers,amsmath,amssymb]{revtex4}
\usepackage{dcolumn}
\usepackage{bm}
\usepackage{epsfig}
\begin{document}
\title{Matter Wave Interference Pattern in the collision of bright solitons (Bose Einstein condensates) in a time dependent trap}
\author{V. Ramesh Kumar$^1$}
\author{R. Radha$^1$}
\email{radha_ramaswamy@yahoo.com}
\author{Prasanta K. Panigrahi$^2$}
\email{prasanta@prl.res.in}
 \affiliation{$^1$ Centre for Nonlinear Science, Department of Physics, Government College for Women, Kumbakonam 612001, India \\
  $^2$ Indian Institute of Science Education and Research, Kolkatta-700106, India}
\begin{abstract}
We show that it is possible to observe matter wave interference
patterns in the collision of bright solitons (Bose Einstein
condensates) without free ballistic expansion for suitable choices
of scattering length and time dependent trap.
\end{abstract}

\maketitle
\section{Introduction}
When a gas of massive bosons are cooled to a temperature very
close to absolute zero in an external potential, a large fraction
of the atoms collapse into the lowest quantum state of the
external potential forming a condensate known as a Bose Einstein
condensate (BEC) [1-4]. Bose Einstein condensation is an exotic
quantum phenomenon observed in dilute atomic gases and has made a
huge turnaround in the fields of atom optics and condensed matter
physics. The recent experimental realization of BECs in rubidium
[5] has really kickstarted the upsurge in this area of research
leading to flurry of activities in this direction while the
observation of dark [6] and bright solitons [7], periodic waves
[8], vortices and necklaces [9] has given an impetus to the
investigation of this singular state of matter.

The dynamics of BECs is in general governed by an inhomogeneous
nonlinear Schrodinger (NLS) equation called the Gross-Pitaevskii
(GP) equation and the behaviour of the condensates depends on the
scattering length (binary interatomic interaction) and the
trapping potential. Eventhough the GP equation is in general
nonintegrable, it has been recently investigated for specific
choices of scattering lengths and trapping potentials [10-12].

It is known that a BEC comprises of coherent matter waves
analogous to coherent laser pulses and the wave nature of each
atom is precisely in phase with that of every other atom. In other
words, the atoms occupy the same volume of space, move at
identical speeds; scatter light of the same color and so on.
Hence, it looked fundamentally impossible to distinguish them by
any measurement. This quantum degeneracy arising out of high
degree of coherence has been recently exploited in an experiment
by Andrews et al. and  Ketterle's group [13] showing that when two
seperate coluds of BECs overlap under free ballistic expansion,
the result is a fringe pattern of alternating constructive and
destructive interference just as it occurs with two interesting
laser radiations. In other words, when BECs were made to collide
upon release from the trap, de Broglie wave interference pattern
containing stripes of high and low density were clearly observed.
These experiments which underlined the high degree of spatial
coherence of BECs led to the creation of atom laser [14]. Can one
observe the same matter wave interference pattern by allowing the
bright solitons which are condensates themselves to collide in a
trap? Motivated by this consideration, we investigate the
collisional dynamics of condensates in a time dependent trapping
potential.

\section{Gross-Pitaevskii equation and Lax-pair}
At the mean field level, the time evolution of macroscopic
wavefunction of BECs is governed by the Gross-Pitaevskii (GP)
equation,
\begin{equation}
i\hbar \frac{\partial \psi (\vec{r},t)}{\partial
t}=\left(\frac{-\hbar^2 }{2m}\nabla^2+g|\psi(\vec{r},t)|^2 + V
\right)\psi(\vec{r},t)
\end{equation}
where $g=4 \pi \hbar^2 a_s (t)/m$, $V=V_0 + V_1$, $V_0(x,y)=m
\omega^2_\perp (x^2+y^2)$, $V_1=m \omega_0^2(t) z^2 / 2$. In the
above equation, $V_0$ and $V_1$ represent atoms in a cylindrical
trap and time dependent trap along z-direction respectively. The
time dependent trap could be either confining or expulsive while
the time dependent atomic scattering length $a_s (t)$ can be
either attractive ($a_s < 0$)or repulsive ($a_s>0$). As a result,
the condensates confront with both time dependent scattering
length and time dependent trapping potential.

Considering BECs as an assembly of weakly interacting atomic
gases, eq. (1) can be effectively reduced to a one dimensional GP
equation taking the following form (in dimensionless units),
\begin{equation}
i\frac{\partial \psi}{\partial
t}+\frac{1}{2}\frac{\partial^2\psi}{\partial z^2}+\gamma (t)
|\psi|^2 \psi -\frac{M(t)}{2}z^2 \psi = 0
\end{equation}
where $\gamma (t) = -2 a_s (t) / a_B$, $M(t)=\omega_0^2 (t)/
\omega^2_\perp$, $a_B$ is the Bohr radius, M(t) describes time
dependent harmonic trap which can be either attractive ($M(t) >
0$) or expulsive ($M(t) < 0 $). Now, to generate bright solitons
of eq. (2) for both regular and expulsive potentials, we introduce
the following modified lens transformation
\begin{equation}
\psi(z,t)=\sqrt{A(t)} Q(z,t) \rm exp(i \Phi(z,t))
\end{equation}
where the phase has the following simple quadratic form
\begin{equation}
\Phi(z,t) = - \frac{1}{2}c(t) z^2.
\end{equation}

Substituting the modified lens transformation given by eq. (3) in
eq. (2), we obtain the modified NLS equation
\begin{equation}
iQ_{t}+\frac{1}{2}Q_{zz}-ic(t)zQ_{z}-ic(t)Q+\gamma(t)A(t)|Q|^{2}Q=0,
\end{equation}
with
\begin{equation}
M(t) = c'(t) - c(t)^2,
\end{equation}
and
\begin{equation}
c(t) = -\frac{d}{dt}\rm ln A(t).
\end{equation}

Equation (5) admits the following linear eigen value problem
\begin{eqnarray}
\phi_z &=& U \phi,\qquad U=\left(
                           \begin{array}{cc}
                             i \zeta(t) & Q \\
                             -Q^* & -i \zeta (t) \\
                           \end{array}
                         \right)\\
\phi_t &=& V \phi,\nonumber\\
                     V &=& \left(\begin{array}{cc}
                                  - i \zeta(t)^2+ i c(t) z \zeta(t) & (c(t)z-\zeta(t))Q \\
                                  + \frac{i}{2}\gamma (t)A(t)|Q|^2 &  + \frac{i}{2}Q_z  \\\\
                                  -(c(t)z-\zeta(t))Q^* &  i \zeta(t)^2-i c(t) z \zeta(t) \\
                                  +\frac{i}{2}Q^*_z  & - \frac{i}{2}\gamma (t) A(t) |Q|^2 \\
                                \end{array}
                              \right)
\end{eqnarray}

In the above linear eigenvalue problem, the spectral parameter
which is complex is nonisospectral obeying the following equation
\begin{equation}
\zeta'(t) = c(t) \zeta (t)
\end{equation}
with $\gamma(t) = 1/A(t)$. It is obvious that the compatability
condition  $(\phi_z)_t = (\phi_t )_z$ generates eq. (5).

Substituting eq. (7) with $\gamma(t) = 1/A(t)$ in eq. (6), we get
\begin{equation}
\gamma ''(t)\gamma(t)-2\gamma'(t)^2-M(t)\gamma(t)^2=0
\end{equation}

Thus, the solvability of the GP eqn.(2) depends on the suitable
choices of scattering length $\gamma(t)$ and the trap frequency
M(t) consitent with eq. (11).  Recently eq. (2) has been
investigated and bright solitons have been generated employing
gauge transformation approach [11].

\section{Bright soliton interaction and matter wave interference}
To investigate the collisional dynamics of condensates in the
presence of a time dependent trap, we now consider the two bright
soliton of the following form,
\begin{equation}
\psi^{2} (z,t) =\sqrt{\frac{1}{\gamma (t)}}
\frac{A_1+A_2+A_3+A_4}{B_1+B_2}\;\;e^{-\frac{i}{2} c(t) z^2}
\end{equation}
where
\begin{eqnarray}
A_1&=& \{-2\beta _2 [(\alpha_2 - \alpha_1 )^2  - (\beta_1^2  -
\beta_2^2 )]\nonumber\\
 &-& 4i\beta_1 \beta_2 (\alpha_2  - \alpha_1 )\}
e^{(\theta_1  + i\xi_2 )}\nonumber
\end{eqnarray}
\begin{equation}
A_2=- 2\beta_2 [(\alpha_2  - \alpha_1 )^2  + (\beta_1^2  +
\beta_2^2 )]e^{( - \theta_1  + i\xi_2 )}\nonumber
\end{equation}
\begin{eqnarray}
A_3&=&\{- 2\beta_1 [(\alpha_2  - \alpha_1 )^2  + (\beta_1^2  - \beta_2^2 )] \nonumber\\
&+& 4i\beta_1 \beta_2 (\alpha_2  - \alpha_1 )\} e^{(i\xi_1  +
\theta_2 )}\nonumber
\end{eqnarray}
\begin{equation}
 A_4=-4i\beta_1 \beta_2 [(\alpha_2  -
\alpha_1 ) - i(\beta_1  - \beta_2 )]e^{(i\xi_1  - \theta_2
)}\nonumber
\end{equation}
\begin{eqnarray}
B_1&=&-4\beta _1 \beta _2 [\sinh (\theta _1 )\sinh (\theta _2 ) +
\cos (\xi _1  - \xi _2)]\nonumber\\
B_2&=& 2\;\cosh (\theta _1 )\;\cosh (\theta _2 )\;[(\alpha _2 -
\alpha _1 )^2  + (\beta _1^2  + \beta _2^2 )] \nonumber
\end{eqnarray}
and
\begin{eqnarray}
\theta_1&=&2  \beta_{1} z-4\int^t_0 (\alpha_{1}\beta_{1})dt^{'}+2\delta_1\nonumber\\
\xi_1&=&2  \alpha_{1} z-2\int^t_0(\alpha_1^2 - \beta_1^2)dt^{'}-2\phi_1\nonumber\\
\theta_2&=&2  \beta_{2} z-4\int^t_0 (\alpha_{2}\beta_{2})dt^{'}+2\delta_2\nonumber\\
\xi_2&=&2  \alpha_{2}z-2\int^t_0(\alpha_2^2 -
\beta_2^2)dt^{'}-2\phi_2\nonumber
\end{eqnarray}
\begin{eqnarray}
\alpha_1&=&\alpha_{1 0}e^{\int^t_0 c(t^{'}) dt^{'}},\qquad
\beta_1=\beta_{10}e^{\int^t_0 c(t^{'}) dt^{'}},\nonumber\\
\alpha_2&=&\alpha_{2 0}e^{\int^t_0 c(t^{'}) dt^{'}},\qquad
\beta_2=\beta_{2 0}e^{\int^t_0 c(t^{'}) dt^{'}},\nonumber\\
\gamma(t)&=&\gamma_0 e^{\int^t_0 c(t^{'}) dt^{'}}.\nonumber
\end{eqnarray}

To generate the interference pattern, we now allow the bright
solitons to collide with each other in the presence of trap for
suitable choices of scattering length $\gamma(t)$ and trap
frequency M(t) (or c(t)) consistent with eq. (11).

\begin{figure}
\epsfig{file=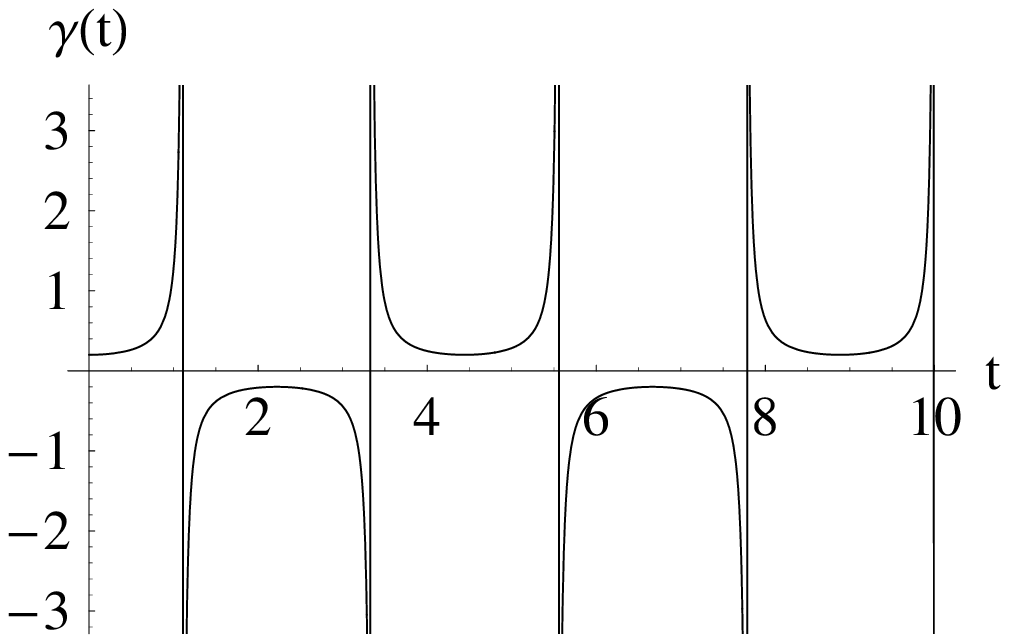, width=0.65\linewidth} \caption{The
variation of scattering length $\gamma[t]$=0.2sec[$\sqrt{2} t$]
corresponding to case (i).}
\end{figure}

\begin{figure}
\epsfig{file=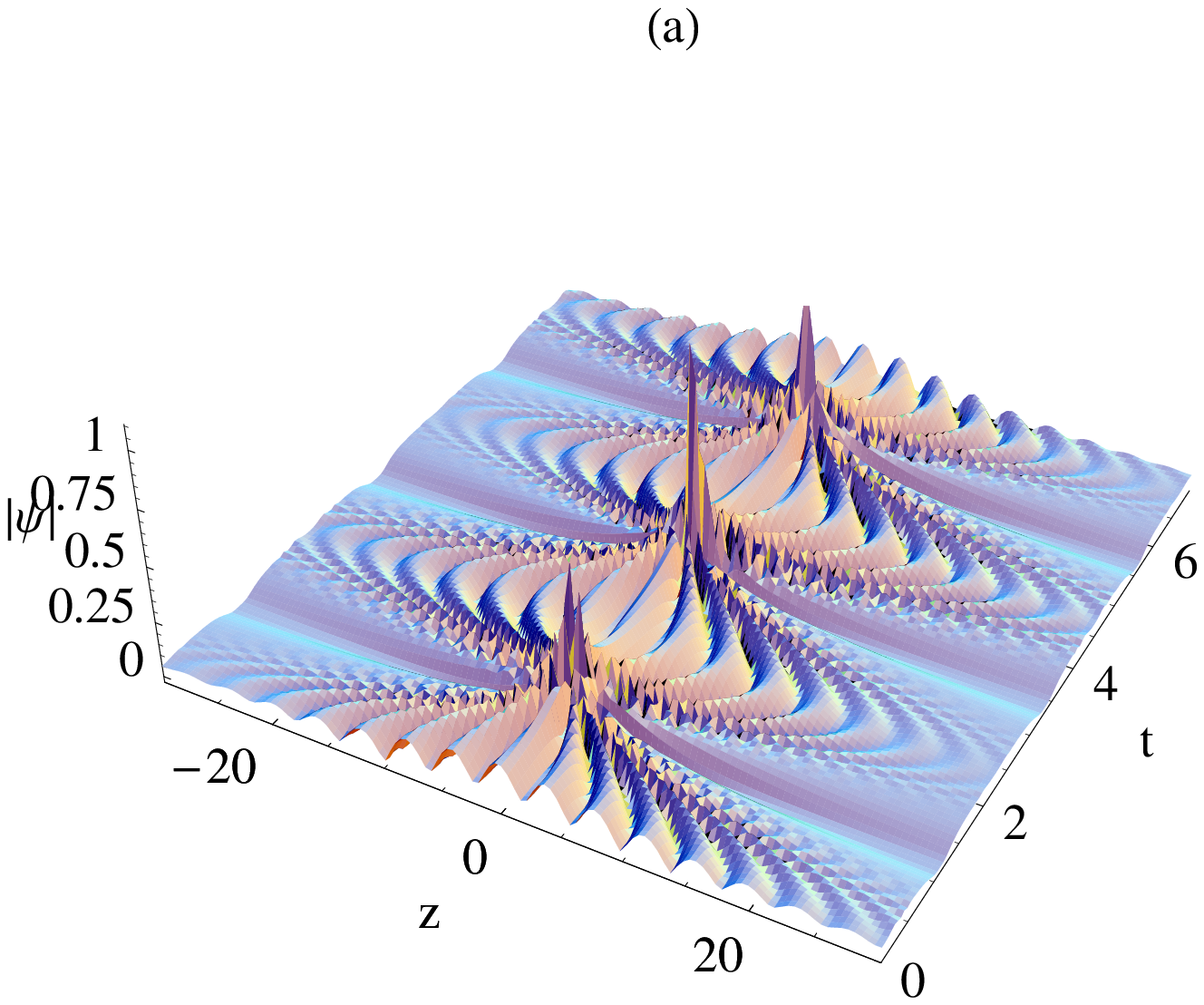, width=0.75\linewidth}
\epsfig{file=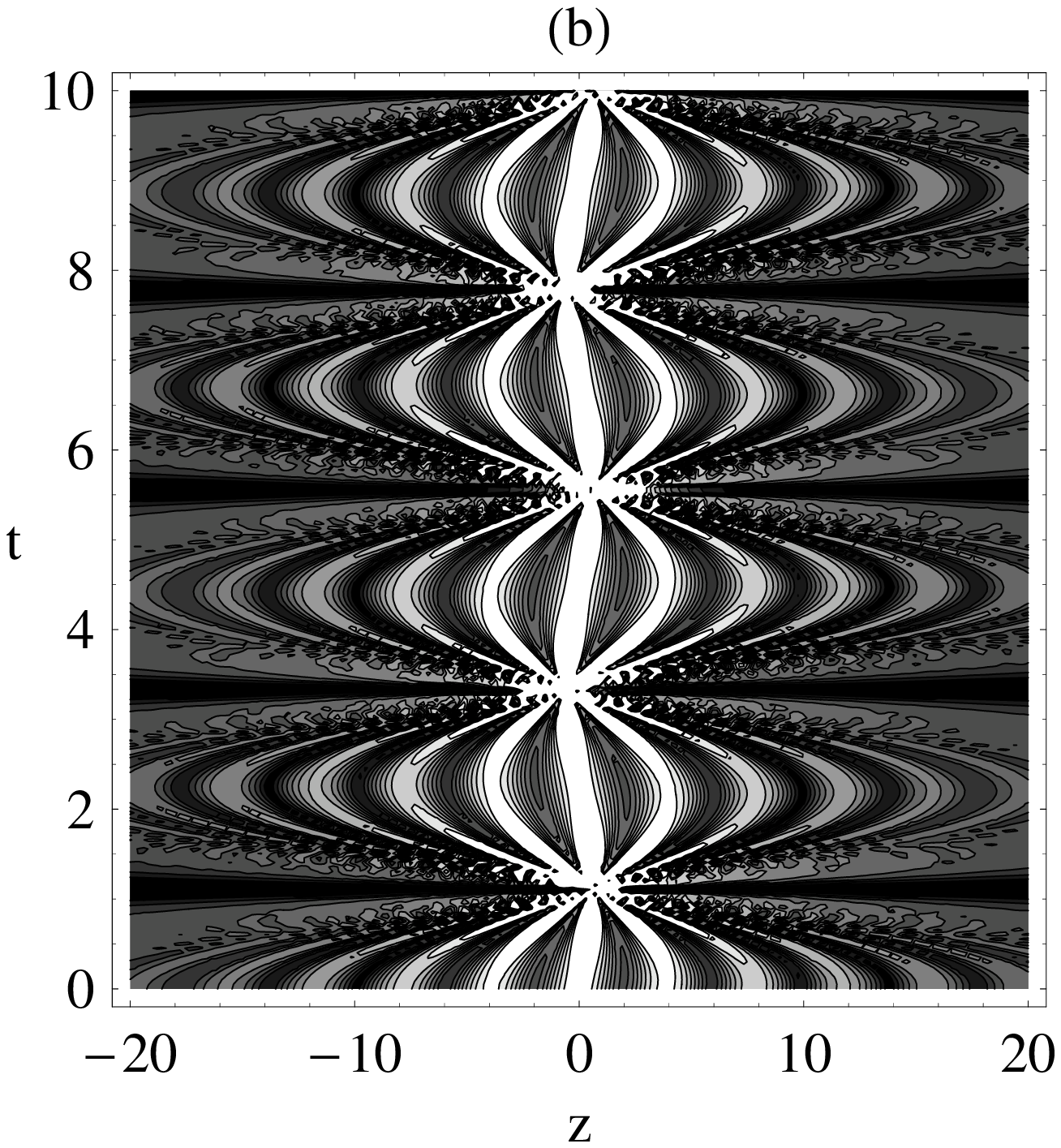, width=0.65\linewidth} \caption{Interaction
of two bright solitons forming interference pattern for the choice
c[t]=$\sqrt{2}$tan[$\sqrt{2}t$], $\gamma[t]$= 0.2sec[$\sqrt{2}
t$], $\alpha_{10}$=.01, $\alpha_{20}$=0.8, $\beta_{10}$=0.06,
$\beta_{20}$=0.012, $\gamma_{0}$=0.2, $\phi_1$ =0.01,
$\phi_2$=0.1, $\delta_1$=0.1, $\delta_2$=0.01 and (b) the
corresponding contour plot depicting matter wave density
evolution}
\end{figure}

\begin{figure}
\epsfig{file=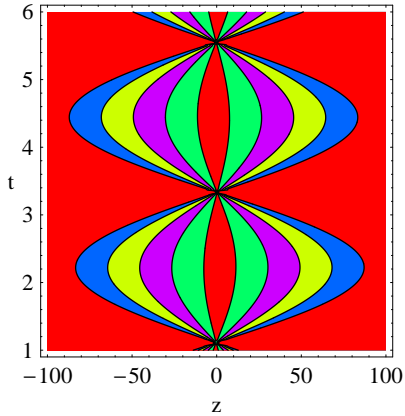, width=0.65\linewidth} \caption{The phase
change evolutions of fig.(2a)}
\end{figure}

Case (i): When c(t)=$\sqrt{2}$tan$\sqrt{2}t$, the variation of
scattering length $\gamma(t)$=0.2sec$\sqrt{2}t$ is shown in fig.1.
Accordingly, the trap frequency M(t) becomes a constant (M(t)=2).
Under this condition, the collisional dynamics of bright solitons
in the harmonic trap is shown in fig.2a and the corresponding
density evolution in fig. 2b. The density evolution consists of
alternating bright and dark fringes of high and low density
respectively while the phase difference between the condensates
shown in fig. 3 continuously changes with time.

Case(ii): When c(t)=$\sqrt{2}$cot$\sqrt{2}t$, the scattering
length varies sinusoidally with time as in the case of Feshbach
resonance (shown in fig.4) while the trap becomes
M(t)=$-2cot^2\sqrt{2}t-2cosec^2\sqrt{2}t$.  The evolution of the
condensates in the above time dependent trap and the corresponding
density evolution are now shown in figs 5a and 5b. From the
interference pattern, one observes that the matter wave density is
maximum at the origin and it decreases gradually on either sides.
Again, the phase change between the condensates oscillates with
time as shown in fig.6.

\begin{figure}
\epsfig{file=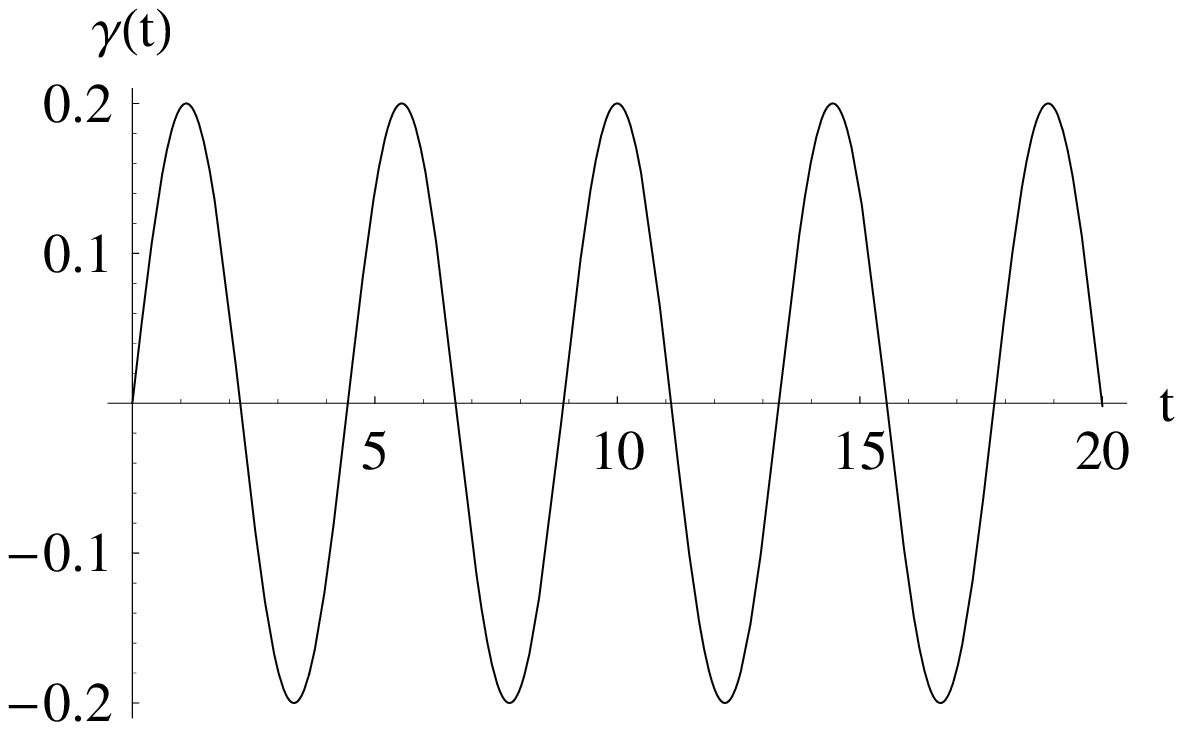, width=0.65\linewidth} \caption{The
variation of scattering length $\gamma[t]$=0.2sin[$\sqrt{2}t$]
corresponding to case (ii).}
\end{figure}
\begin{figure}
\epsfig{file=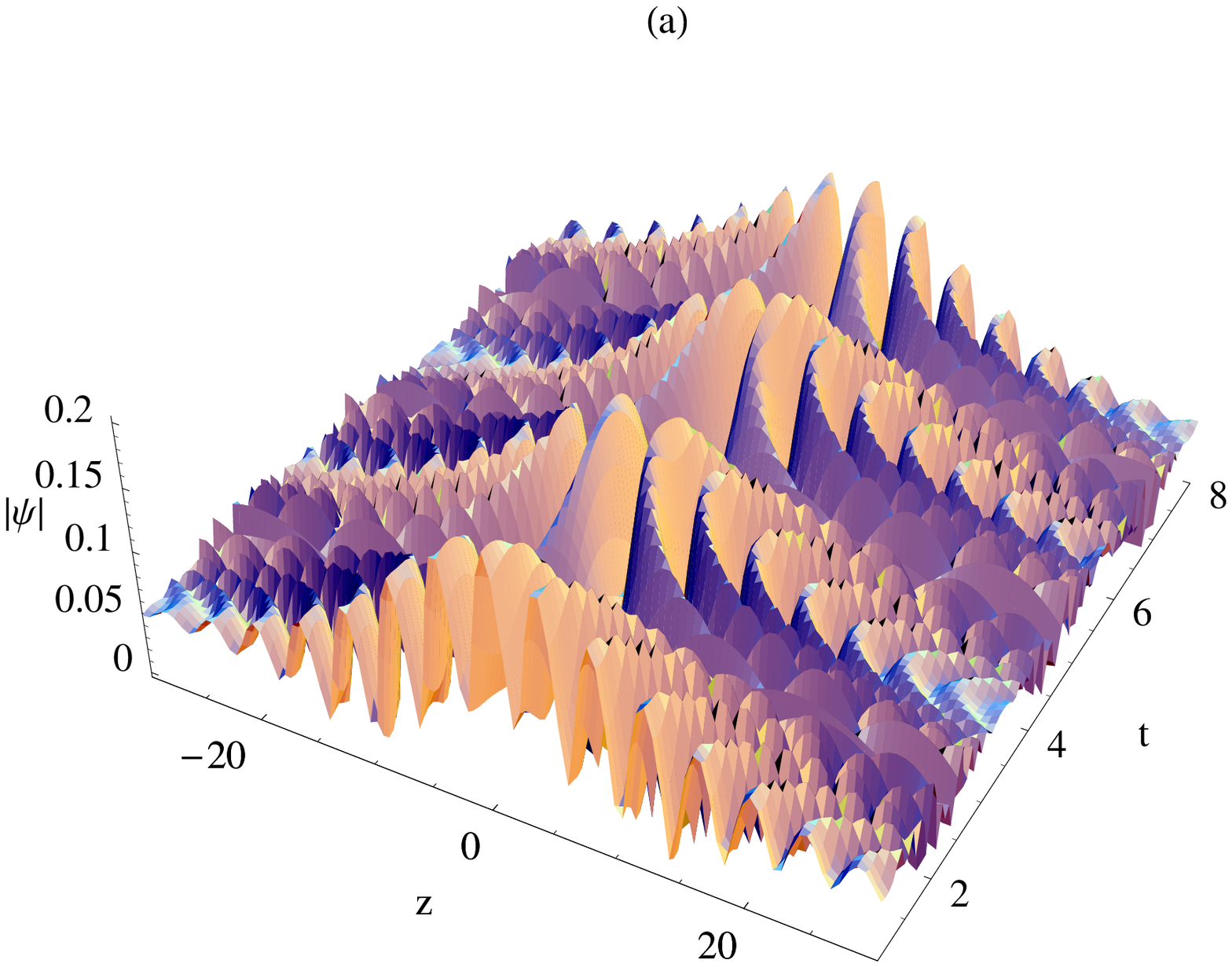, width=0.75\linewidth}
\epsfig{file=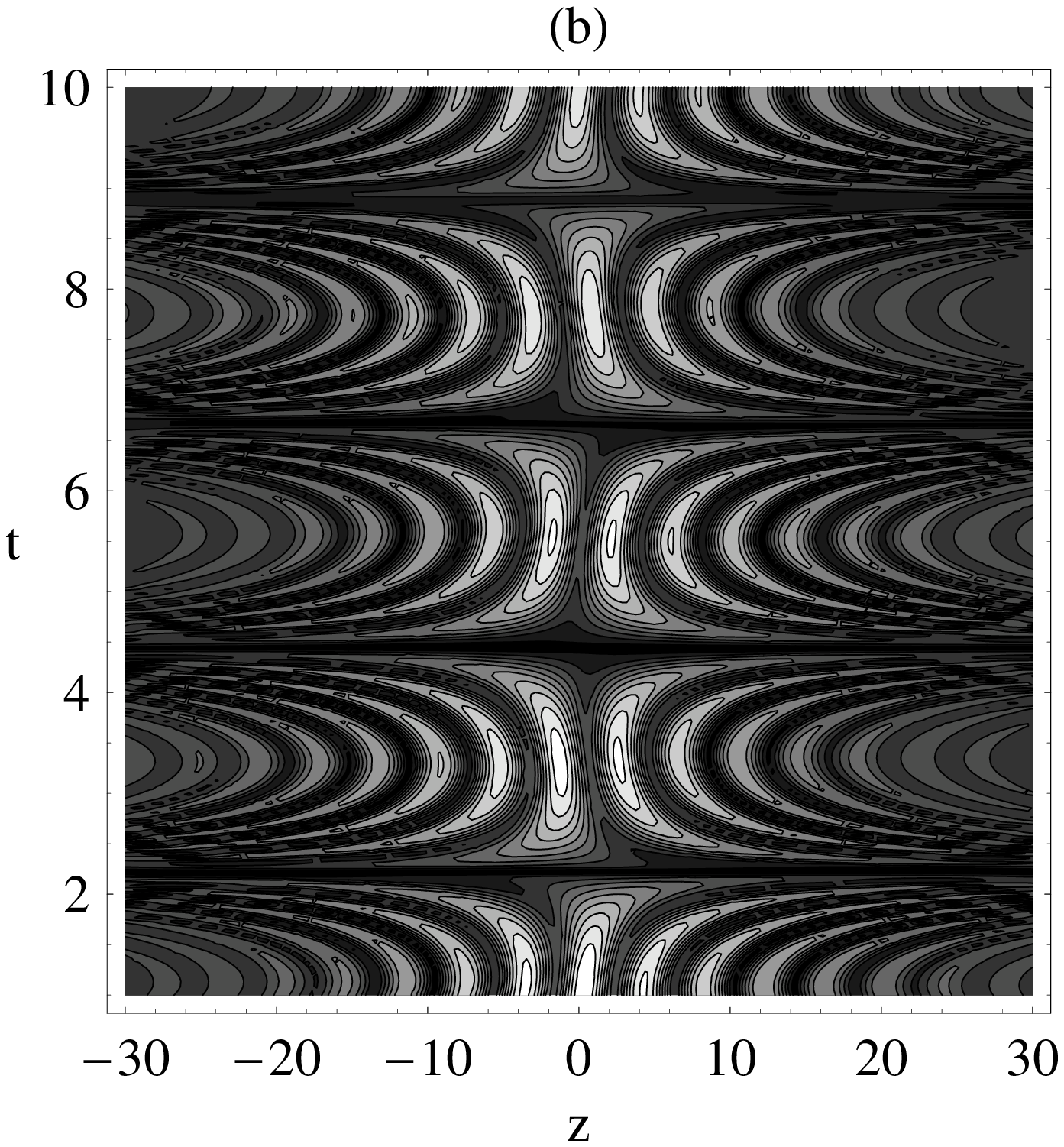, width=0.65\linewidth} \caption{Two soliton
interaction forming interference pattern for the choice
c[t]=$\sqrt{2}$cot[$\sqrt{2}t$], $\gamma[t]$=0.2 sin[$\sqrt{2}t$]
with the other parameters as in fig (1a) and (b) the corresponding
contour plot depicting matter wave density evolution.}
\end{figure}
\begin{figure}
\epsfig{file=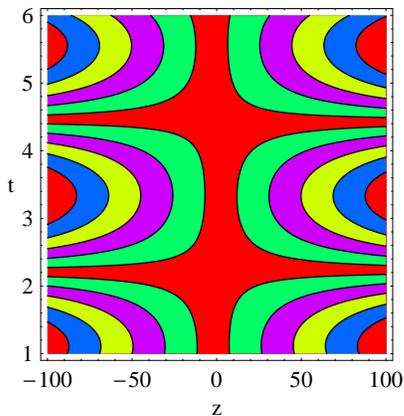, width=0.65\linewidth} \caption{The phase
change evolution of fig. 5a}
\end{figure}

Thus, our results reinforce the fact that the matter waves
originating from the condensates (bright solitons) do interfere
and produce a fringe pattern analogous to the coherent laser beams
and the interference pattern is a clear signature of the long
range spatial coherence of the condensates. It should be mentioned
that the interference patterns were obtained earlier by Andrews et
al. [13] under the condition of free ballistic expansion (atoms
are released from the trap) while we selectively tune the
frequency of the trap M(t) in accordance with the scattering
length $\gamma(t)$ (consistent with eq.(11)) to observe matter
wave interference in the collisional dynamics of bright solitons.
It should be mentioned that the optical traps have opened up the
possibility of realizing different types of temporal variation of
the trap frequency M(t) while the scattering length $\gamma(t)$
can be controlled both by Feshbach resonance as well as through
the trap frequency M(t). The phase change evolution which gives a
measure of the coherence of the condensates shown in figs.3 and 6
shows that it oscillates with time and this is reminiscent of
Josephsen effect wherein the phase difference between the
superconducting wave functions oscillates with time.

\section{Conclusion}
In conclusion, we have shown that it is possible to observe matter
wave interference pattern by studying the collisional dynamics of
condensates (bright solitons) for suitable choices of scattering
length $\gamma(t)$ and trap frequency M(t) and this could happen
without free ballistic expansion. We also observe that the phase
difference between the condensates oscillates with time clearly
showing the manifestation of Josephson effect. It would be
interesting to interpret the matter wave destructive interference
and whether this would mean that atoms plus atoms add up to
vaccuum.

\section{Acknowledgements}
VR wishes to thank DST for offering a Senior Research Fellowship.
RR acknowledges the financial assistance from DST and UGC in the
form of major and minor research projects.

\end{document}